\documentclass[a4paper, 10pt]{article}

\usepackage[latin1]{inputenc}
\usepackage[dvips]{graphicx} 
\usepackage{epsfig}
\usepackage{gensymb}
\usepackage{times, mathptmx}
\usepackage[square, sort&compress]{natbib}
\usepackage{setspace}
\usepackage{moon}
\usepackage{caption}
\usepackage{fancyhdr}

\title{Lunar Satellite Detection of Ultra-High Energy Neutrinos with the Use of Radio Methods}
\author{O. St{\aa}l$^1$, J. Bergman$^1$, B. Thid\'e$^{1,2}$, L. Åhl\'en$^1$ and G. Ingelman$^3$}
\affiliation{$^1$ Swedish Institute of Space Physics (IRF-U)\\
\		  Box 537, SE-751 21 Uppsala, Sweden\\
		  E-mail: os@irfu.se\\
\vspace{6pt}
	     $^2$ LOIS Space Centre, V\"axjö University\\
	          SE-351 95 V\"axj\"o, Sweden\\
\vspace{6pt}	          
	     $^3$ High Energy Physics, Uppsala University\\
\		  Box 535, SE-751 21 Uppsala, Sweden
	      }
\date{}	
\begin{document}
\maketitle
\thispagestyle{fancyplain}

\begin{abstract}
Neutrinos interact with matter only through weak processes with low cross-section. To detect cosmic neutrinos most efforts have relied on the detection of visible Vavilov-\v{C}erenkov light in detectors embedded in the target volumes. To access the decreasing flux of ultra-high energy neutrinos, far above 1 PeV, ideas on how to increase the detection volume by observing coherent radio frequency emission caused by the Askaryan effect have been put forward. Here we describe how a satellite in lunar orbit equipped with 
an electromagnetic vector sensor could detect Askaryan pulses induced by neutrinos interacting with the moon. The threshold neutrino energy is found to be $5\times10^{19}$ eV for this setup, and the sensitivity is determined from simulations. A model dependent event rate of $2.2$ events per year is calculated.
\end{abstract}

\begin{spacing}{1}
\section*{Introduction}
The high end of the cosmic neutrino energy spectrum remains unstaked territory. Different models (Active Galactic Nuclei \citep{AGN}, Topological Defects \citep{TD}, GZK-cutoff \citep{GZK1, GZK2}) predict sources of cosmic neutrinos with energy far above 1 PeV (10$^{15}$ eV), termed \emph{ultra-high energy} (UHE) neutrinos. Validation of these models remains an experimental challenge.

Detection of cosmic neutrinos is possible by observing optical Vavilov-\v{C}erenkov emission from secondary particle showers produced in dense media from neutrino interactions. With the expected decrease of the neutrino flux with energy \citep{Gandhi00}, larger detector volumes are necessary for higher energy particles. The ICECUBE detector will equip $1$ km$^3$ \citep{ICECUBE} of South pole ice with photomultiplier tubes and still larger volumes will be needed to access the UHE neutrinos. It becomes increasingly impractical to detect the UHE neutrinos by their induced optical radiation, also since the optical output is generated in an incoherent process and therefore the output power increases only linearly with the neutrino energy.

One way to overcome this limitation and access a sufficient detector volume is to detect Vavilov-\v{C}erenkov radio emissions. At radio frequencies, the emission is generated coherently and the output power is proportional to the \emph{square} of the primary energy. Coherent emission is possible since the showers develop a negative charge excess going through the material; this is known as the \emph{Askaryan effect} \citep{Askaryan62}, \citep{Askaryan65}. 

Radio methods for neutrino detection have been employed by several groups world-wide. One such experiment was performed by the GLUE collaboration, who utilised the Goldstone radio telescopes in search for pulses from the Moon \citep{GLUE}. Embedded aerials in the Antarctic ice have been used by the RICE \citep{RICE} group. Data from the FORTE satellite \citep{FORTE} has been searched for pulses generated in the Greenlandic ice sheet. No events were reported from any of these experiments, and this sets upper limits on the neutrino flux above 10$^{17}$ eV. Of great promise for the future is ANITA \citep{ANITA}, a balloon-carried transient detector to be launched in 2006.

In this report, we will discuss how a digital broadband vector radio receiver on board a satellite orbiting the Moon could be used to detect radio pulses from neutrinos interacting with the lunar surface. To detect Askaryan pulses from the Moon by satellite measurements has previously been suggested by \textit{Gorham} \citep{Gorham04}.

\section*{Radio emission from particle showers}
Neutrinos interact with matter only through weak processes with low cross-sections. The cross-section increases slowly with energy, and determines a minimum neutrino energy required for interaction with the lunar material. This energy can be estimated by comparing the energy dependent interaction length, given by the cross-section, to the lunar radius $R_\mathrm{L} = 1738$ km. Using an average density of $3.34$ g/cm$^3$ for the lunar interior \citep{NSS} gives a value of $E_\mathrm{min}\simeq 3\times 10^{16}$ eV. Above this energy, the Moon is opaque to neutrinos.

After the first interaction, a shower of secondary elementary particles is generated. The charged component consists mainly of a compact region of electrons and positrons propagating at superluminal speeds \citep{Leroy99}. Perpendicular to the direction of motion, the spatial scale of the shower is given by the Molière radius \citep{PDG}, which for Si is about 5 cm \citep{Leroy99}. On the average, 90\% of the shower energy is contained inside a cylinder of this radius around the shower core.

\subsection*{Radiation in homogeneous media}
For a point charge moving with uniform velocity in a homogeneous, dielectric medium of refractive index $n=\sqrt{\epsilon}$, we can obtain, in the standard way, the Fourier transform of the electric field by direct solution of the Maxwell's equations \citep{Thide04}. Considering satellite detection, Fraunhofer geometry applies. The vectorial distance traversed by the charge is denoted $\mathbf{d}$, with a component $\mathbf{d_{\perp}}$ perpendicular to the direction to the observation point. Following \citep{Zas92}, the radiated electric field becomes
\begin{equation}
\label{fullE}
\mathbf{E}_{\omega}(\mathbf{x})=\frac{\mathrm{i}\omega \mu q}{4\pi \epsilon_0 c^2}\frac{e^{\mathrm{i}kr}}{r}e^{\mathrm{i}(\omega-\mathbf{k}\cdot \mathbf{v})t_1}\mathbf{d_{\perp}}\, e^{\mathrm{i}\phi}\left[\frac{\sin{\phi}}{\phi}\right]
\end{equation}
where the square-bracketed expression resembles the angular dependence of a slit-diffraction pattern. An overall phase is provided by $t_1$, the time when the particle starts to radiate. The phase-factor
\begin{equation}
\label{phase}
\phi=\frac{\pi d}{\lambda}\left(1-n\beta \cos{\theta}\right)\\ ,
\end{equation}
is defined in terms of $d=|\mathbf{d}|$, the observation angle $\theta$ from the shower axis, and the velocity factor $\beta$ = $|\mathbf{v}|/c$. If the particle velocity exceeds the phase speed of light in the medium $v=c/n$, there is a maximum field strength attained for $\phi=0$. This condition defines the \v{C}erenkov angle $\theta_\mathrm{c}=\arccos{1/(n\beta)}$. When the phase factor is small, a McLaurin expansion of the phase dependence in (\ref{fullE}) is possible. Retaining only the first term, the field expression simplifies to
\begin{equation}
\label{approxE}
\mathbf{E}_{\omega}(\mathbf{x})\approx \frac{\mathrm{i}\omega \mu q}{4\pi \epsilon_0 c^2}\frac{e^{\mathrm{i}kr}}{r}e^{\mathrm{i}(\omega-\mathbf{k}\cdot \mathbf{v})t_1}\mathbf{d_{\perp}}\ .
\end{equation}
The necessary criterion for (\ref{approxE}) to be valid is either that the wavelength is sufficiently long ($d \ll \lambda$) or that the observation angle is close to the \v{C}erenkov angle. 

The total field from a shower is obtained by adding the contributions of all particles. The result depends on the spatial scales, as compared to the wavelength of the radiation. In the optical regime, all particles radiate incoherently and the individual phases must be taken into account through (\ref{fullE}). The individual phase factors are uniformly distributed which makes the signs of the charges unimportant and the field strength scales with the square root of the primary energy. 

The radio regime is at the other end of the spectrum. If all spatial scales of the shower are assumed to be smaller than the wavelength, the radiation will be emitted almost coherently and the field strength will scale linearly with energy. In this regime,  Eq. (\ref{approxE}) is valid, and only the net charge gives rise to the field. 

Between the two regimes there is a transition regime of partial coherence. Closer to the \v{C}erenkov angle, higher frequency components display coherence. Following \citep{Zas92}, we define the highest frequency with coherent emission at the \v{C}erenkov angle as the \emph{specific decoherence frequency} of the material.

\subsection*{The development of charge asymmetry}
For a shower with equal numbers of electrons and positrons, no coherent radiation can be emitted. Askaryan realized that when such a neutral shower develops in a material, scattering processes will create an imbalance with a negative charge excess \citep{Askaryan62}. The most important contribution comes from Compton scattering $(\gamma + e_{\mathrm{atom}}^- \rightarrow \gamma + e^-)$ which incorporates new electrons into the shower. Another process increasing the number of shower electrons is Bhaba scattering $(e^+ + e_{\mathrm{atom}}^- \rightarrow e^+ + e^-)$. Finally, there is annihilation $(e^+ + e_{\mathrm{atom}}^- \rightarrow \gamma + \gamma)$, further reducing the positron component. Typical values for the fully developed charge excess are about 20\% \citep{Zas92}.

\subsection*{Spectral parametrization}
The total radio pulse spectrum has been determined from simulations taking into account the primary neutrino interaction and the following shower development. Such simulations for showers in ice \citep{Zas92} of total energy $W_0$ have been generalized to other dense media and can be summarized in the empirical formula \citep{Muniz97}
\begin{equation}
\label{spectrum}
|\mathbf{E_{\omega}}(r, \theta_\mathrm{c})|=\frac{A}{r}\left(\frac{W_0}{1\  \mathrm{TeV}}\right)\frac{\omega}{\omega_0}\frac{1}{1+0.4\left(\frac{\omega}{\omega_0}\right)^{1.44}}\ \ \mathrm{[Vm^{-1}MHz^{-1}]}\  .
\end{equation}
The amplitude $A$ is determined empirically, as is the specific decoherence frequency $\omega_0$. The general features of this formula have been verified experimentally at SLAC \citep{SLAC}. With bunches of GeV photons released in a Si target, a total shower energy of 10$^{19}$ eV was attained. We will adopt the experimental parameters for Si to the radiation from the Moon, namely $A = 2.53\cdot10^{-7}$ V MHz$^{-1}$ and $\omega_0 = 2.5$ GHz. 

Showers from primaries at an energy above 1 PeV interacting initially through the electromagnetic channel are known to be elongated by the Landau-Pomeranchuk-Migdal (LPM) effect \citep{Landau52, Migdal56}. This significantly reduces the angular spread of the Vavilov-\v{C}erenkov cone \citep{Muniz97}, making such showers unfavourable for detection. They will therefore not be considered in the following. Neutrinos interacting initially through the hadronic channel will not suffer LPM elongation since many particles of lower energy are generated at an early stage of the shower development. An average fraction 0.2 of the primary energy is deposited in the hadronic shower component \citep{Alvarez98}.

The coherent radiation from showers of many particles is not limited to the \v{C}erenkov angle, but, as discussed above, for low frequencies the coherence is maintained further off-axis. The reduction in field strength off-centre approximately follows a Gaussian:
\begin{equation}
\label{spread}
|\mathbf{E_{\omega}}(r, \theta)|=|\mathbf{E_{\omega}}(r, \theta_{\mathrm{c}})|e^{-\frac{1}{2}\left(\frac{\theta-\theta_{\mathrm{c}}}{\Delta \theta}\right)^2}
\end{equation}
with a frequency dependent width given by $\Delta \theta \approx 2.4\degree \omega_0/\omega$.

\section*{The Moon as a neutrino detector}
The uppermost layer of the lunar surface consists of a fine-grained material known as the \emph{regolith}. The electrically neutral regolith is composed of single mineral grains, small rock fragments and combinations of these, fused by impacts. The depth of the regolith is believed to be at least $2$--$8$ metres in the maria, increasing to 15 meters or more in highland regions \citep{NSS}. Silicates are the most abundant component in the regolith, but important for the dielectric characteristics are also the varying concentrations of Fe and Ti oxides.

Assuming the mass density of the regolith layer to be constant $\rho = 1.7$ g/cm$^3$ over the depths considered ($0$--$10$ m), the dielectric constant is taken to have an average value of $\epsilon = n^2=3$. It has been found previously that the value of the loss tangent depends most significantly on the concentrations of TiO$_2$ and FeO in the material. Assuming an average total concentration of these metallic oxides of 5\%, the loss tangent $\tan{\delta}=0.003$ \citep{Olhoeft75}. This gives an attenuation length of 100 m for a 100 MHz signal. The attenuation length is inversely proportional to the frequency. 

\section*{Satellite detection of Askaryan radio pulses}
\subsection*{Threshold}
From the empirical formula (\ref{spectrum}) it is possible to estimate the threshold energy for detection, assuming the background level is known. The required signal level is commonly given as a power figure in dBm per Hz, related to the spectral field strength through
\begin{equation}
P (\mathrm{dBm}) = 10\log_{10}\left(\frac{P}{1\ \mathrm{mW\cdot Hz^{-1}}}\right)\ ,
\end{equation}
\begin{equation}
E_\omega = \frac{10^6}{h_\mathrm{A}}\sqrt{\frac{P\, R}{\Delta\omega}}\ \ \ \mathrm{[\mu V\, m^{-1}\, MHz^{-1}]}
\end{equation}
where $h_\mathrm{A}$ is the effective antenna height, $R=50$ $\Omega$ and $\Delta\omega$ the system bandwidth. Equating with (\ref{spectrum}) now gives the threshold energy
\begin{equation}
\label{th}
W^\mathrm{min}_0 = 2\times 10^{25}\left(\frac{r}{h_\mathrm{A}}\right)\left(\frac{\omega_0}{\omega}\right)\sqrt{\frac{P\, R}{\Delta\omega}}\ \ \ \mathrm{[eV]}\ .
\end{equation}
This analysis is valid in the regime where $E_\omega$ increases linearly with $\omega$ [cf. Eq (4)], \textit{i.e.} for $\omega + \Delta\omega \ll \omega_0 $.

\subsection*{Aperture and event rate}
In order to estimate the expected event rate, a calculation of the effective lunar aperture (area $\times$ solid angle) for neutrino detection as a function of energy is required. The expected event rate is then obtained by convolving the differential flux with the aperture
\begin{equation}
\frac{dN_\mathrm{\nu}}{dt}=\int_{W_\mathrm{min}}^{\infty}\mathrm{d}W_\mathrm{\nu}\frac{\partial^2N_\mathrm{\nu}}{\partial t \partial W_\mathrm{\nu}}[A\Omega]_\mathrm{geom}\, S(W_\mathrm{\nu})\ .
\end{equation}
Where the aperture $A\Omega$ is factorized as the geometric aperture $[A\Omega]_\mathrm{geom}$ multiplying a specific efficiency, $S$. We determine this efficiency by Monte Carlo simulations.

Since the angle for total internal reflection at the Moon-vacuum interface $\theta_{\mathrm{TIR}}$ is complementary to $\theta_\mathrm{c}$, only emission from showers that are upgoing with respect to the surface will escape when the \v{C}erenkov cone is narrow. For lower frequencies the cone is wider, and total internal reflection is not a significant problem since the transmission rises to a maximum just a few degrees off $\theta_\mathrm{TIR}$. 
\begin{figure}
\centering
\includegraphics[width=8cm]{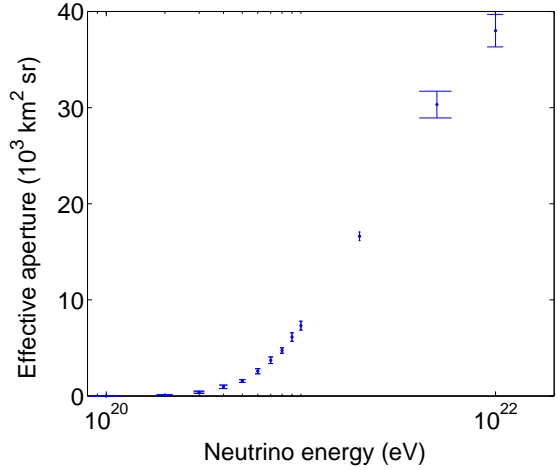}
\caption{Effective aperture for lunar neutrino detection. One sigma error bars indicate statistical deviations.}
\end{figure}

All rays will be refracted at the interface. For a planar surface, the refraction will follow Snell's law. This will broaden the \v{C}erenkov peak and the acceptance angle is increased on the expense of the field strength. A more elaborate simulation of the interface transmission should take the structure of the surface into account.

\subsection*{Experimental equipment}
The ELVIS (ELectromagnetic Vector Information Sensor) instrument \citep{ELVIS} recently proposed for the Indian Moon mission Chandrayaan-1, scheduled for launch in 2007--2008, provides a suitable test-bed for detection of lunar neutrinos by radio methods. The equipment consists of three 5 m dipole aerials in an orthogonal configuration, providing full 3D sampling of the electric field vector and the advantage of polarimetric measurements. The signals are digitized in linear 14 bit ADCs. The sampling frequency is currently limited to 80--125 Msamples/s, but when operated in transient mode, signals up to 200--500 MHz are detectable.

\begin{figure}
\centering
\includegraphics[width=8cm]{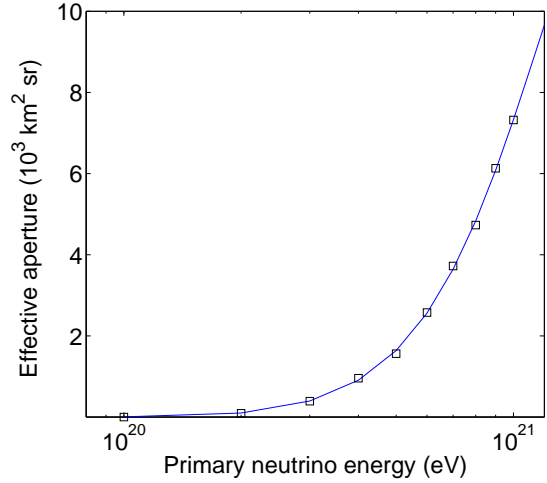}
\caption{Aperture in the interesting energy regime close to threshold with a polynomial fit to the simulation results.}
\end{figure}
\subsection*{Simulation results}
Calculations of the aperture and expected event rate has been performed for the ELVIS and Chandrayaan-1 setup using our own Monte Carlo code. The detector parameters used were 100 km height above the surface,  $\omega = 100$ MHz for the observation frequency with an estimated bandwidth of $50$ MHz. The minimum detectable signal strength was taken as a conservative $-135$ dBm. The threshold according to formula (\ref{th}) becomes $5\times 10^{19}$ eV, but, as can be seen from the results, the aperture below 10$^{20}$ eV is effectively negligible. 

The aperture was calculated using cross-sections from \textit{Gandhi et al.} \citep{Gandhi98}, ZHS parametrization of the field according to (\ref{fullE}), full ray-tracing and smooth surface interactions as outlined above. The result is presented in Figure 1, showing a complicated energy dependence near threshold followed by a logarithmic increase of aperture with primary energy. The region close to threshold is presented more clearly in Figure 2, with a polynomial fit used for event rate integration.

Adapting the GZK neutrino flux of \textit{Engel et al.} \citep{Engel01} as a standard neutrino source at ultra-high energy, the integrated event rate over a year of observation time becomes $2.2$ expected events for these parameter values.

\section*{Conclusions and outlook}
Radio detection of coherent Vavilov-\v{C}erenkov emission originating from showers induced in the moon by ultra-high energy cosmic neutrinos has been reviewed. Calculations for a satellite detection experiment reveals the very low event rate to expect, but it should be possible to improve the event rate by orders of magnitude through optimization of the experimental parameters.  

Highly interesting is also to study the use of an antenna array on the lunar surface for neutrino detection. This must be performed through extensive simulations. Such simulations are underway, giving further results and more insights in the near future.

\end{spacing}
\bibliographystyle{prsty}
\bibliography{lurefs}

\end{document}